\documentclass{andp2012}
\usepackage[english]{babel}
\usepackage{graphicx}
\usepackage{subfig}
 
\usepackage{bm}
\usepackage{mathdots}

\newcommand{\Tr}{\mathrm{Tr}}
\title{Charging of quantum batteries with general harmonic power}
\author{Jie Chen\inst{1}}
\author{Liyao Zhan\inst{1}}
\author{Lei Shao\inst{1}}
\author{Xingyu Zhang\inst{1}}
\author{Yuyu Zhang\inst{2}}
\author{Xiaoguang Wang\inst{1,} \footnote{E-mail:~\textsf{xgwang1208@zju.edu.cn}}}
\address[1]{Zhejiang Institute of Modern Physics, Department of Physics, Zhejiang University, Hangzhou 310027, China}
\address[2]{Department of Physics, Chongqing University, Chongqing 401330, China}
\shortauthors{J. Chen et al.}

\begin{abstract}
To explore further improvement of the charging efficiency in the quantum batteries system, the charging process with general harmonic driving field is studied in this paper. The charge saturation is introduced to describe the charging efficiency, with which the charging mode is divided into the saturated charging mode and the unsaturated charging mode. The relationships between the time-dependent charge saturation and the parameters of general driving field are discussed both analytically and numerically. And the expressions of time-dependent charge saturation with the quasienergy and the Floquet states of system is given with the Floquet theorem. With both analytical and numerical results, the optimal parameters to reach the best charging efficiency are found.
\end{abstract}
\shortabstract
\begin{document}
\maketitle
\section{Introduction}\label{sec1}
With the development of quantum thermodynamics and the growing demand for device miniaturization in recent years, quantum batteries (QBs), as quantum energy storage devices, have gradually become a focus of research in quantum physics~\cite{QB2013inPhysRevE.87.042123,QB2015byBinder_2015,QB2018inPhysRevA.97.022106,QB2017inPhysRevLett.118.150601,QB2018inPhysRevLett.120.117702,QB2018inPhysRevB.98.205423,QB2018byfriis2018precision,QB2019-2byZhang,QB2019inPhysRevB.99.035421,QB2019inPhysRevLett.122.210601,QB2019inPhysRevB.99.205437,QB2019inPhysRevLett.122.047702}. The QBs' energy storage process is also well-known as the charging process, which is to transfer the state of the system from lower energy levels to higher energy levels, and the charging system usually consists of the QBs system $H_{0}$ and the charging field $V\left(t\right)$, as $H\left(t\right)=H_{0}+V\left(t\right)$~\cite{QB2013inPhysRevE.87.042123,QB2015byBinder_2015}. With the power of charging field, the state of charging system transfers from the initial state $\rho_{0}=\rho\left(0\right)$ (on lower energy levels) to the final state $\rho\left(t\right)$ (on higher energy levels), and the stored energy of QBs during the process is~\cite{QB2013inPhysRevE.87.042123,QB2017inPhysRevLett.118.150601}
\begin{eqnarray}
E\left(t\right) & = & \Tr\left[\left(\rho\left(t\right)-\rho\left(0\right)\right)H_{0}\right].\label{eq:Et}
\end{eqnarray}

\begin{figure}
\centering \includegraphics[scale=0.22]{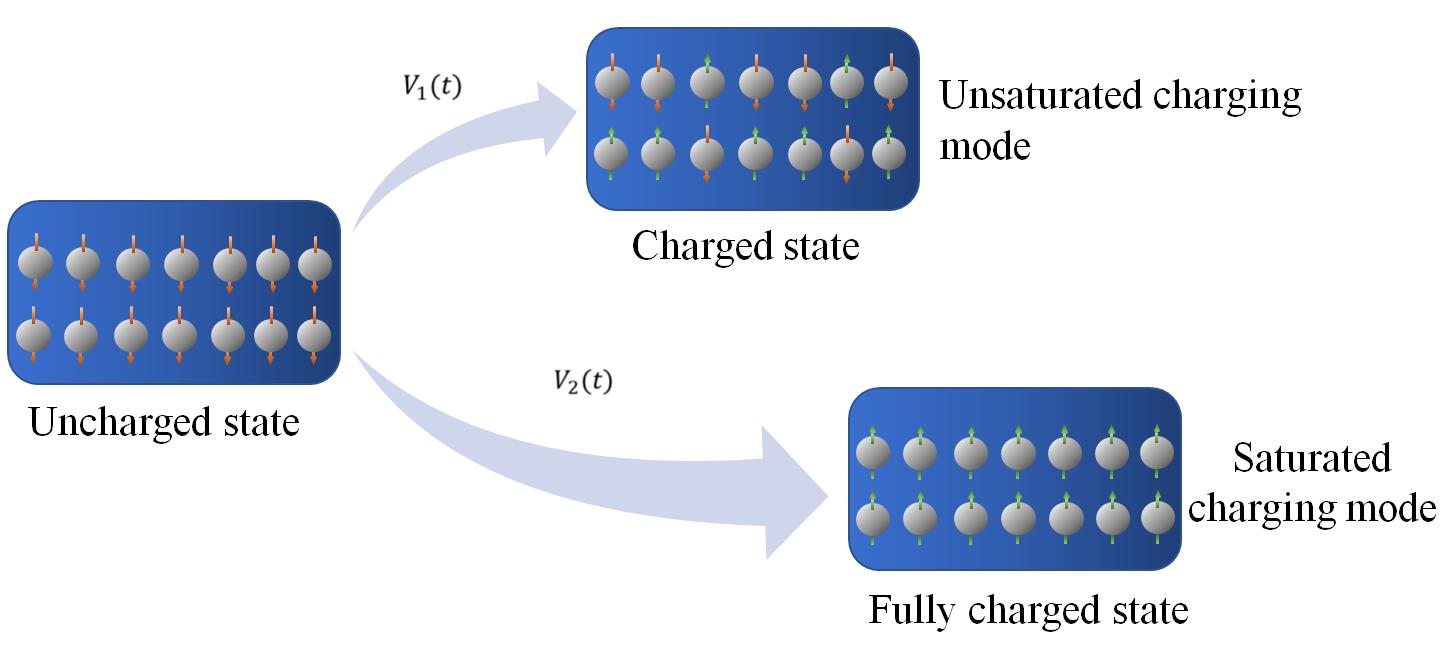}\protect\caption{The type of charging fields determines the charge saturation and the charging power of QBs system. With different charging field, the QBs will be charged in saturated or unsaturated charging mode. Here, we choose a general harmonic charging field as $V\left(t\right)=\sum_{i=x,y,z}A_{i}\cos\left(\omega_{i}t+\phi_{i}\right)J_{i}$,
and find out the optimal harmonic charging field by adjusting the parameters $A_{i}$, $\omega_{i}$ and $\phi_{i}$, $i=x,y,z$.}
\end{figure}

The results of time-dependent stored energy $E\left(t\right)$ can describe the charging process perfectly, and how to improve the charging efficiency in the charging process is a key issue of research on QBs~\cite{QB2013inPhysRevE.87.042123,QB2015byBinder_2015,QB2018inPhysRevA.97.022106,QB2017inPhysRevLett.118.150601}. Some previous studies have shown that choosing a proper type of charging fields plays an important role in improving the charging efficiency~\cite{CAinPhysRevLett.111.240401,QB2017inPhysRevLett.118.150601,QB2018inPhysRevLett.120.117702,QB2018byfriis2018precision,QB2019-2byZhang}.
For example, charging with a harmonic driving field is more efficient than charging with a steady one~\cite{QB2018inPhysRevA.97.022106,QB2019-2byZhang}, and when the charging process contains entangling operations, it will attain a collective quantum advantage, i.e., the more the number of battery units, the faster the charging speed~\cite{CAinPhysRevLett.111.240401,QB2017inPhysRevLett.118.150601,QB2018inPhysRevLett.120.117702}. For further explanation of charging efficiency, we introduce two more concepts based on the energy-time relation $E\left(t\right)$ as:\\
\textbf{Charge saturation}, which is defined as the ratio between the charged energy in the process and the fully charged energy of battery pack system;\\
\textbf{Charging power}, which is well-known as the charging speed, and defined as the ratio between the charged energy and the charging time~\cite{QB2015byBinder_2015,QB2017inPhysRevLett.118.150601,QB2018inPhysRevA.97.022106,QB2018inPhysRevLett.120.117702}.

When we do research on the time-dependent charge saturation of QBs with different charging fields, we find that the charge saturation can never reach $100\%$ with some kinds of charging fields. But, the unsaturated QBs are also able to provide enough energy with plenty of charged battery units. In this way, the effective charging mode of QBs can be divided into the saturated charging mode and the unsaturated charging mode. Then after setting a proper threshold of charge saturation as the minimum effective charge saturation, the charging power can be directly reflected by the charging time which is spent to reach the threshold value. So the efficiency of the charging process can be directly reflected with the time-dependent charging saturation.

The system of quantum battery (QB) unit can be simplified as a two-level system since the energy level is discrete in quantum system, and it can be described by the charged state
$\left(\begin{array}{c}1\\0
\end{array}\right)$
 and uncharged state
$\left(\begin{array}{c}0\\1
\end{array}\right)$~\cite{QB2018inPhysRevA.97.022106,QB2019-2byZhang}.
 To supply adequate energy supplement, battery pack system is usually constituted by many battery units. Then the Hamiltonian of QBs can be written as $H_{0}=\sum_{i}\frac{1}{2}\omega_{0}^{\left(i\right)}\sigma_{z}^{i}$ (setting $\hbar=1$), where $\omega_{0}^{\left(i\right)}$ is the stored energy of single QB unit when it is charged. In this paper, the QB units are set to be identical, i.e., $\omega_{0}^{\left(i\right)}=\omega_{0}$, so we have $H_{0}=\frac{1}{2}\omega_{0}\sum_{i=1}^{N}\sigma_{i}^{z}=\omega_{0}J_{z}$~\cite{QB2018inPhysRevA.97.022106,QB2019-2byZhang}. Then we choose the general harmonic charging field as $V\left(t\right)=\sum_{i=x,y,z}A_{i}\cos\left(\omega_{i}t+\phi_{i}\right)J_{i}$,
where $A_{i},\omega_{i},\phi_{i},i=x,y,z$ are the driving strengths, driving frequencies and initial phases in the three orthogonal directions, respectively. So the Hamiltonian of the whole charging system is written as
\begin{eqnarray}
H\left(t\right) & = & \omega_{0}J_{z}+\sum_{i=x,y,z}A_{i}\cos\left(\omega_{i}t+\phi_{i}\right)J_{i}.\label{eq:Ht}
\end{eqnarray}
It is a quantum driven system which can be described by the Dicke states $|s,m\rangle(m=-s,-s+1,\cdots,s)$, i.e. the eigenstates of $J^{2}$ and $J_{z}$. And compared with the semi-classical Rabi model, it can be named as semi-classical Dicke model with generalized driving fields. In this paper, we study the relationship between the time-dependent charge saturation and the parameters of the driving fields since they are all adjustable. The relationships will help find out the optimal parameters to reach the best charging efficiency.

Obviously, the fully charged energy of our QBs system is $N\omega_{0}$. So the time-dependent charge saturation is written as~\cite{QB2015byBinder_2015,QB2017inPhysRevLett.118.150601,QB2018inPhysRevA.97.022106,QB2018inPhysRevLett.120.117702}
\begin{eqnarray}
\eta\left(t\right) = \frac{E\left(t\right)}{N\omega_{0}}.\label{eq:eta_t}
\end{eqnarray}

Of course, some other factors also should be taken into consideration since they are crucial for enhancing the practicality of QBs, such as the ratio between the stored energy of QBs system and the input energy provided by charging field. In this paper, we introduce $A=|\bm{A}|^{2}=\sum_{i=x,y,z}A_{i}^{2}$ to control the input energy and rewrite the strengths of three directions as
$\bm{A}=\left(A_{x},A_{y},A_{z}\right)=(A\cos\Theta\cos\Phi,  A\cos\Theta\sin\Phi,$ $A\sin\Theta )$, where $A=|\bm{A}|$ and $\Theta\in[-\frac{\pi}{2},\frac{\pi}{2})$,
$\Phi\in[0,2\pi)$.

However, with the general harmonic charging field, the system is difficult to be solved analytically. In previous research, a kind of analytically solvable situation with $\Theta=0,\Phi=0,\phi_{x}=0$ has been studied by Yuyu Zhang, et al. in Ref~\cite{QB2019-2byZhang}. In this paper, we find some more analytical solutions with the charging in one or two directions. Then we compare charging efficiencies of these situations in saturated charging mode and unsaturated charging mode, respectively.

Furthermore, according to the Floquet theorem, the charging system can be described by the Floquet Hamiltonian since it is periodic~\cite{Floquet1inPhysRev.138.B979,Floquet1byCHU20041,Floquet2009inPhysRevA.79.032301,CHRWAandFloquetinPhysRevA.96.033802}.
And with the eigenvalues and eigenstates of the Floquet Hamiltonian, which are also named as the quasienergy and Floquet states of the periodic system, the time-dependent charge saturation of the system can be expressed analytically.

In this paper, we first analyze some kinds of analytically solvable QBs charging systems, including reviewing the results of the system given in former work~\cite{QB2019-2byZhang}. With the results, we make a comparison between the charging systems on the charging efficiency. Furthermore, under the framework of Floquet theorem, we give the analytical expressions of time-dependent charge saturation with the quasienergy and the Floquet states of the system. Then after the
solutions with analytical methods, we also discuss the relationship between the charging efficiencies and variety of parameters in charging field $V\left(t\right)$ with some numerical results. And in this way, we find the optimal parameters to reach the best charging efficiency. Finally, we give our conclusions and some further discussions.

\section{Charging efficiencies of QBs system with charging in one or two directions}\label{sec2}

As uncharged state, the initial state of the system is  $|\psi\left(0\right)\rangle=\left(\begin{array}{c}
0\\
1
\end{array}\right)^{\otimes N}$, which can be rewritten as $|\psi\left(0\right)\rangle=|\frac{N}{2},-\frac{N}{2}\rangle$ with the representation of Dicke states, and the initial density matrix can be gotten as $\rho\left(0\right)=|\psi\left(0\right)\rangle\langle\psi\left(0\right)|$.

\subsection{Analytical results with charging in parallel direction}\label{sec2-1}

Firstly, we consider about the charging in parallel direction. By setting $\Theta=\frac{\pi}{2}$, the Hamiltonian of our charging system turns into $H\left(t\right)=\omega_{0}J_{z}+A\cos\left(\omega t+\phi\right)J_{z}$, which describes the system merely with a driving in the parallel direction. With the Schr\"{o}dinger equation, the final state of QBs system is easy to get as
\begin{eqnarray}
|\psi\left(t\right)\rangle & = & \exp\left\{ i\frac{2k\pi}{\sin\phi}\left[\frac{\omega_{0}\omega}{A}t+\sin\left(\omega t+\phi\right)\right]\right\} |\frac{N}{2},-\frac{N}{2}\rangle,
\end{eqnarray}
where $k\in\mathbb{Z}$ is an arbitrary integer. The results tell that the QBs system cannot be charged with the harmonic charging field in the parallel direction, since there is only the change of phase on the state.

\subsection{Analytical results with charging in one vertical direction}\label{sec2-2}
Then when the QBs are charged with the harmonic power in one vertical direction, Hamiltonian of the charging system is written as $H\left(t\right)=\omega_{0}J_{z}+A\cos\left(\omega t+\phi\right)J_{\bot}$, where $J_{\bot}=\cos\Phi J_{x}+\sin\Phi J_{y}$ represents the action in arbitrary vertical direction. With $U_{z}=e^{i\Phi J_{z}}$, as the unitary transformation in su(2) system~(see Appendix A for details), the Hamiltonian could be simplified as
\begin{eqnarray}
\tilde{H} \left(t\right) = \omega_{0}J_{z}+A\cos\left(\omega t+\phi\right)J_{x},\label{eq:tildeH}
\end{eqnarray}
with $|\tilde{\psi}\left(t\right)\rangle = e^{i\Phi J_{z}}|\psi\left(t\right)\rangle$. And it is easy to prove that the charge saturation remains unchanged after the transformation as $\tilde{\eta}\left(t\right)=\langle\psi\left(t\right)|e^{-i\Phi J_{z}}\frac{1}{N}J_{z}e^{i\Phi J_{z}}|\psi\left(t\right)\rangle+\frac{1}{2}=\langle\psi\left(t\right)|\frac{1}{N}J_{z}|\psi\left(t\right)\rangle+\frac{1}{2}=\eta\left(t\right)$.
So we use the Hamiltonian in Eq.(\ref{eq:tildeH}) (named as $\tilde{H}$-system briefly) to describe arbitrary charging process with the harmonic power in one vertical direction.

To solve the system, we need to extend the counterrotating hybridized rotating wave approximation (CHRWA) from two-level system to su(2) system~\cite{Scully2000QuantumOptics,CHRWA2012inPhysRevA.86.023831,CHRWA2015inPhysRevA.91.053834,CHRWAbyCJ}. At first step, the unitary transformation $\tilde{U}_{1} \left( t \right)=\exp\left[ i\frac{A}{\omega}\xi\sin\left(\omega t+\phi\right)J_{x}\right]$
is introduced with $\xi$ as a regulating variable, and the Hamiltonian turns to be
\begin{small}
\begin{eqnarray}
\tilde{H}_{1} &\! =\! & \omega_{0}\!\left\{ \cos\!\left[\frac{A}{\omega}\xi\sin\left(\omega t \!+\!\phi\right)\right]\!J_{z}\!+\!\sin\!\left[\frac{A}{\omega}\xi\sin\left(\omega t\!+\!\phi\right)\right]\!J_{y}\right\} \nonumber\\
& &+A\left(1-\xi\right)\cos\left(\omega t+\phi\right)J_{x}.
\end{eqnarray}
\end{small}

Then the $\tilde{H}_{1}$ can be expanded as infinite series with $\cos\left(z\sin\phi_t\right)=J_{0}\left(z\right)+\sum_{k=1}^{\infty}2J_{2k}(z)\cos\left(2k\phi_t\right)$ and $\sin\left(z\sin\phi_t\right) =\sum_{k=0}^{\infty}2J_{2k+1}(z)\sin\left[\left(2k+1\right)\phi_t\right]$, where $J_{n}(\cdot)$ is the Bessel function. And after neglecting the higher-order harmonic terms as the approximation, the Hamiltonian is rewritten as~\cite{CHRWA2012inPhysRevA.86.023831,CHRWA2015inPhysRevA.91.053834,CHRWAbyCJ}
\begin{eqnarray}
\tilde{H}_{1}  & = &  \omega_{0}J_{0}\left(\frac{A}{\omega}\xi\right)J_{z}+A\left(1-\xi\right)\cos\left(\omega t+\phi\right)J_{x}\nonumber\\
& &+2\omega_{0}J_{1}\left(\frac{A}{\omega}\xi\right)\sin\left(\omega t+\phi\right)J_{y},\nonumber\\
 & = & \omega_{0}J_{0}\left(\frac{A}{\omega}\xi\right)J_{z}\!+\!\tilde{A}\left[\cos\left(\omega t \!+\!\phi\right)J_{x}\!+\!\sin\left(\omega t\!+\!\phi\right)J_{y}\right],
\end{eqnarray}
where $\tilde{A}=A\left(1-\xi\right)=2\omega_{0}J_{1}\left(\frac{A}{\omega}\xi\right)$ is regulated by $\xi$. Then with the transformation
$\tilde{U}_{2}\left(t\right)=\exp\left[i\left(\omega t+\phi\right)J_{z}\right]$, the Hamiltonian turn into time-independent, as
\begin{eqnarray}
\tilde{H}_{2} & = & \tilde{\Delta}J_{z}+\tilde{A}J_{x},
\end{eqnarray}
where $\tilde{\Delta}=\omega_{0}J_{0}\left(\frac{A}{\omega}\xi\right)-\omega$ is the effective detuning~\cite{CHRWA2012inPhysRevA.86.023831,CHRWA2015inPhysRevA.91.053834,CHRWAbyCJ}.

Furthermore, with $\tilde{U}_{3}\left(t\right)=\exp\left(-i\tilde{H}_{2}t\right)$ as the time evolution operator of the $\tilde{H}_{2}$-system, the time-dependent state $|\tilde{\psi}_{2}\left(t\right)\rangle$ can be expressed as $|\tilde{\psi}_{2}\left(t\right)\rangle=\tilde{U}_{3}\left(t\right)|\tilde{\psi}_{2}\left(0\right)\rangle$. And with $|\tilde{\psi}_{2}\left(t\right)\rangle=\tilde{U}_{2}\left(t\right)\tilde{U}_{1}\left(t\right)|\tilde{\psi}\left(t\right)\rangle$, the initial state $|\tilde{\psi}_{2}\left(0\right)\rangle$ is gotten as $|\tilde{\psi}_{2}\left(0\right)\rangle=\tilde{U}_{2}\left(0\right)\tilde{U}_{1}\left(0\right)|\tilde{\psi}\left(0\right)\rangle$ by setting $t=0$. So the state of $\tilde{H}$-system can be written approximately as
\begin{eqnarray}
|\tilde{\psi}\left(t\right)\rangle=\tilde{U}_{1}^{\dagger}\left(t\right)\tilde{U}_{2}^{\dagger}\left(t\right)\tilde{U}_{3}\left(t\right)\tilde{U}_{2}\left(0\right)\tilde{U}_{1}\left(0\right)|\frac{N}{2},-\frac{N}{2}\rangle.
\end{eqnarray}

Especially, when the initial phase $\phi=0$, the expression of state turns to be $|\tilde{\psi}\left(t\right)\rangle=\tilde{U}_{1}^{\dagger}\left(t\right)\tilde{U}_{2}^{\dagger}\left(t\right)\tilde{U}_{3}\left(t\right)|\frac{N}{2},-\frac{N}{2}\rangle$, and it is the system which has been studied~\cite{QB2019-2byZhang}.
After substituting the state into Eq.~(\ref{eq:Et}) and Eq.~(\ref{eq:eta_t}), the time-dependent charge saturation of $\tilde{H}$-system can be written as
\begin{eqnarray}
\tilde{\eta}\left(t\right)& = &\frac{1}{2}\left[1\!-\!\cos\tilde{\phi}\!+\!\frac{\sin\left(\Omega_{R}t\right)}{\Omega_{R}}\tilde{A}\cos\left(\omega t\right)\sin\tilde{\phi}\right.\nonumber\\
& &\left.+\frac{\cos\left(\Omega_{R}t\right)\!-\!1}{\Omega_{R}^{2}}\left(\tilde{A}\tilde{\Delta}\sin\left(\omega t\right)\sin\tilde{\phi}\!-\!\tilde{A}^{2}\cos\tilde{\phi}\right)\right],\label{eq:tildeeta}
\end{eqnarray}
where $\tilde{\phi}=\frac{A}{\omega}\xi\sin\left(\omega t\right)$, $\Omega_{R}=\sqrt{\tilde{\Delta}^{2}+\tilde{A}^{2}}$ , and $\xi$ is the root of regulating formula  $A\left(1-\xi\right)=2\omega_{0}J_{1}\left(\frac{A}{\omega}\xi\right)$.

\begin{figure}[ht]
	\centering 
	\includegraphics[width=90mm]{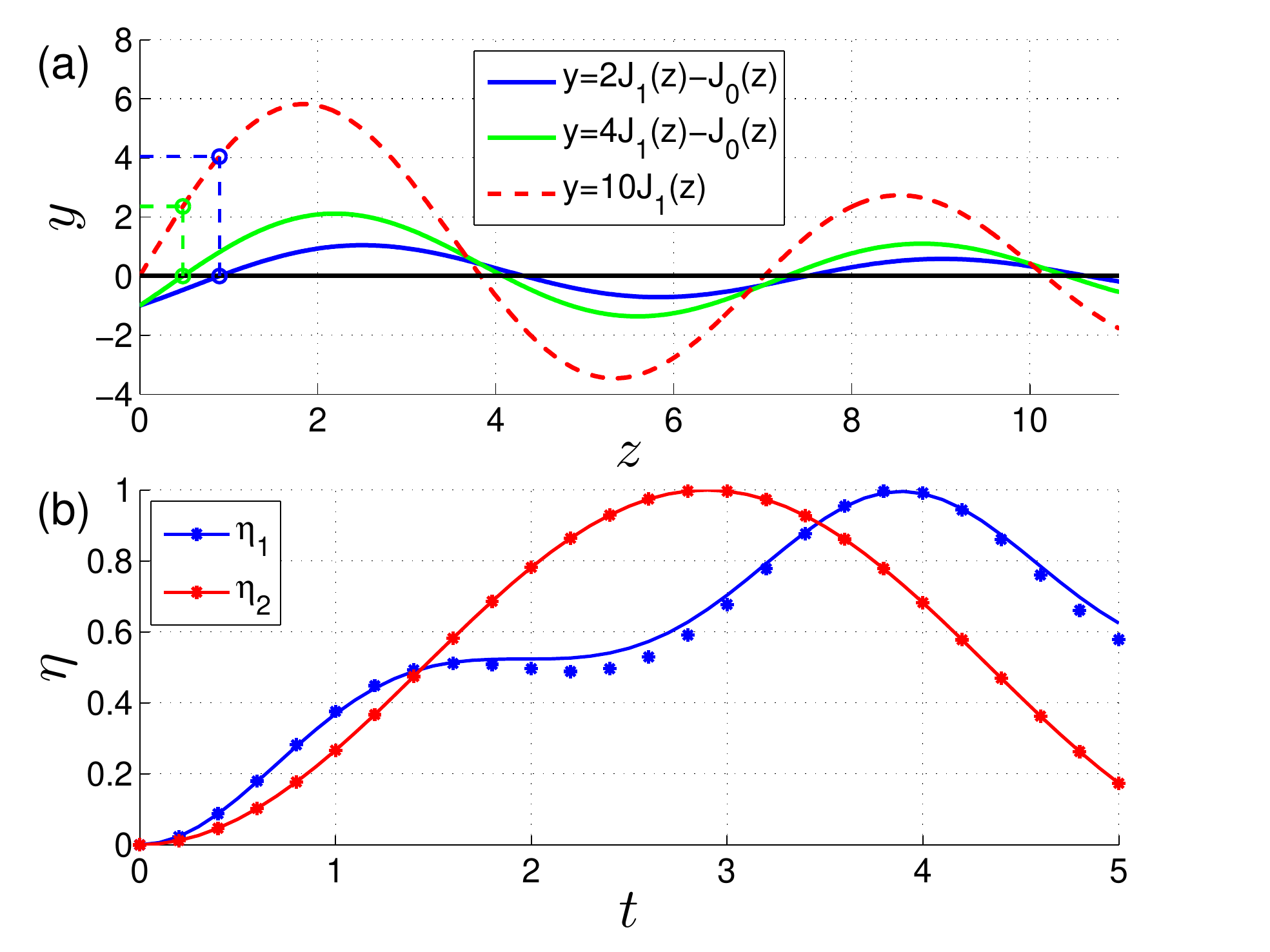}
	\protect
	\caption{(a) Graphic method to solve equation $J_{0}\left(z\right)=kJ_{1}\left(z\right)$ and the comparison of the roots with the dotted red line $y=10J_1\left(z\right)=\frac{5\pi}{\omega_{0}}\tilde{t}^{-1}_{\min}$;
		(b) The time-dependent charging saturation of the $\tilde{H}$-system (as $\eta_{1}$, the blue one) and the $H$-system (as $\eta_{2}$, the red one), where the stars and the solid lines stand for the analytical results and the numerical results, respectively.	}
	\label{fig:Fig2}
\end{figure}

Eq.~(\ref{eq:tildeeta}) tells that we can get $\eta=1$ when $\Omega_{R}=\tilde{A}$, ~$\omega t=k\pi$~ and ~$\Omega_{R} t=(2k'+1)\pi$, with $k,k'\in \mathbb{N}$. So the charging should be stopped at $\tilde{t}=(2k'+1)\pi/\tilde{A}$ and it is obviously that the minimum charging time $\tilde{t}_{\min}$ is gotten when $k'=1$. Furthermore, by introducing $z=\dfrac{A}{\omega}\xi$, the parameter conditions to reach maximum charging efficiency can be rewritten as $A =\omega_{0}J_{0}\left(z\right)\left(z+\frac{1}{k}\right)$ and $\omega =\omega_{0}J_{0}\left(z\right)$, where $z$ is the root of transcendental equation $J_{0}\left(z\right)=2kJ_{1}\left(z\right)$. We slove the equation with graphic method in Fig.\ref{fig:Fig2}(a), and when $k=1$ and $k=2$, the series of the roots of the equation are marked as the intersection points between the graph of function $y=kJ_{1}\left(z\right)-J_{0}\left(z\right)$ (the solid blue and green line stand for $k=1$ and $k=2$, respectively) and $z$-axis. Then we introduce the dotted red line as $y=10J_1\left(z\right)=\frac{5\pi}{\omega_{0}}\tilde{t}^{-1}_{\min}$ to compare the roots and find that the first positive root with $k=1$ is optimal. So is it when compared to the situation of $k>2$. It means that when we set charging strength and charging frequency of the $\tilde{H}$-system as
\begin{eqnarray}
\begin{cases}
A & =\omega_{0}J_{0}\left(z\right)\left(z+1\right),\\
\omega & =\omega_{0}J_{0}\left(z\right),
\end{cases}
\end{eqnarray}
the QBs are fully charged at the time
\begin{eqnarray}
\tilde{t}_{\min}=\frac{\pi}{J_{0}\left(z\right)}\omega_{0}^{-1},
\end{eqnarray}
where $z \approx 0.90$ is the first positive root of transcendental equation $J_{0}\left(z\right)=2J_{1}\left(z\right)$.

\subsection{Analytical results with charging in two vertical directions}\label{sec2-2}

\begin{figure}[htbp]
	\centering \includegraphics[width=90mm]{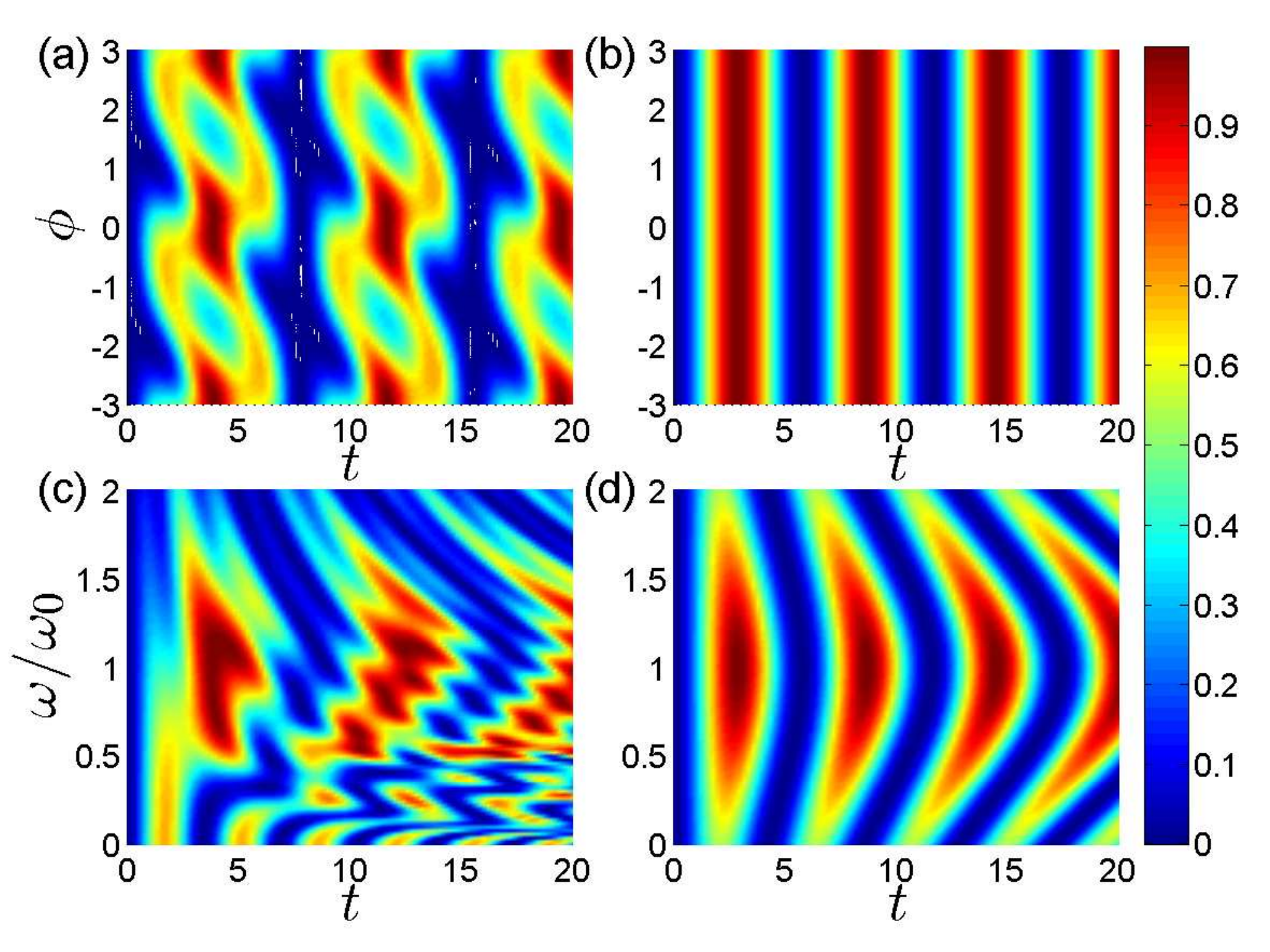}
	\protect\caption{(a) Relationship between the charge saturation and the initial phase $\phi$ in $\tilde{H}$-system, as  $\eta_{1}\left(\phi,t\right)$, when $A=1.53\omega_{0}$ and $\omega=0.81\omega_{0}$ \textbf{(the charging time $t$ takes $\omega_{0}^{-1}$ as unit for all figures in this paper)}; (b) Relationship between the charge saturation and the initial phase $\phi$ in $H$-system, as $\eta_{2}\left(\phi,t\right)$, when $A=1.53\omega_{0}$ and $\omega=\omega_{0}$;
		(c) Relationship between the charge saturation and the charging frequency $\omega$ in $\tilde{H}$-system, as $\eta_{1}\left(\omega,t\right)$, when $A=1.53\omega_{0}$ and $\phi=0$; (d) Relationship between the charge saturation and the charging frequency $\omega$ in $H$-system, as $\eta_{2}\left(\omega,t\right)$, when $A=1.53\omega_{0}$ and $\phi=0$.}
	\label{fig:Fig3}
\end{figure}

Then we adjust the initial phases and set $\phi_{x}=\phi,\phi_{y}=\phi-\frac{\pi}{2}$ to get a system charged with harmonic power in two vertical directions as
\begin{eqnarray}
H\left(t\right) & = & \omega_{0}J_{z}\!+\!\frac{\sqrt{2}}{2}A\left[\cos\left(\omega t \!+\!\phi\right)J_{x}\!+\!\sin\left(\omega t \!+\!\phi\right)J_{y}\right],\label{eq:H_twomode}
\end{eqnarray}
which is an analytically solvable system (briefly named as $H$-system).

The Hamiltonian is easy to be converted into time-independent with unitary transformation $U_{1}\left(t\right)=\tilde{U}_{2}\left(t\right)=\exp\left[i\left(\omega t+\phi\right)J_{z}\right]$
as
\begin{eqnarray}
H_{1} & = & \left(\omega_{0}-\omega\right)J_{z}+\frac{\sqrt{2}}{2}AJ_{x},
\end{eqnarray}
so does the wave function into $|\psi_{1}\left(t\right)\rangle=U_{1}\left(t\right)|\psi\left(t\right)\rangle$.
Then with $|\psi_{1}\left(t\right)\rangle=e^{-iH_{1}t}|\psi_{1}\left(0\right)\rangle$,
we can get the wave function of $H$-system as
\begin{eqnarray}
|\psi\left(t\right)\rangle & = & U_{1}^{\dagger}\left(t\right)e^{-iH_{1}t}U_{1}\left(0\right)|\frac{N}{2},-\frac{N}{2}\rangle.
\end{eqnarray}

Furthermore, the analytical expression of charge saturation in $H$-system is obtained as
\begin{eqnarray}
\eta\left(t\right) & = & \frac{A^{2}}{4\Omega^{2}}\left[1-\cos\left(\Omega t\right)\right],\label{eq:etaH}
\end{eqnarray}
where $\Omega=\sqrt{\left(\omega_{0}-\omega\right)^{2}+\frac{1}{2}A^{2}}$, and the charge saturation is independent of the initial phase. Eq.~(\ref{eq:etaH}) tells that when $\omega=\omega_{0}$, the $H$-system is fully charged (as $\eta=1$) at $t_{\min}=\frac{\sqrt{2}\pi}{A}$.\\

Then we compare the charging efficiency of $\tilde{H}$-system and $H$-system in the saturated charging mode. In the $\tilde{H}$-system, with $z \approx 0.90$, the charging strength and charging frequency is set as $A \approx 1.53 \omega_{0}$ and $\omega \approx 0.81 \omega_{0}$ in this mode. So the QBs are fully charged with the same charging strength at $\tilde{t}_{\min}\approx 3.88 \omega_{0}^{-1}$ and $t_{\min} \approx 2.90\omega_{0}^{-1}$ in the $\tilde{H}$-system and $H$-system, respectively. The analytical and numerical results are shown by the time-dependent charge saturation of the two systems in Fig.\ref{fig:Fig3}(b). It is obviously that the $H$-system has advantages in saturated charging mode.

For further exploration of relations between the charge saturation and the parameters in the unsaturated charging mode, we set $A=1.53\omega_{0}$ and give the numerical results of charge saturation $\eta$ with the variety of initial phase $\phi$ and driving frequency $\omega$ in Fig.\ref{fig:Fig3}(a)(b) and (c)(d), respectively.

The results in Fig.\ref{fig:Fig3} tells that the charging efficiency of $\tilde{H}$-system can be enhanced in the unsaturated charging mode, but it is still less than the one in $H$-system. In a word, the $H$-system have the advantage on charging QBs both in the saturated charging mode and unsaturated charging mode when compared with the $\tilde{H}$-system.

\section{Expression of charge saturation with Floquet theorem}\label{sec3}

Now we come back to the general harmonic charging field. For arbitrary charging frequencies $\omega_{x,y,z}$, we could find an $\omega$ to set $\omega_{i}=n_{i}\omega$, $i=x, y, z$, where $n_{x,y,z}$ are positive integers. So Hamiltonian of the charging system is periodical with $T=\frac{2\pi}{\omega}$, which means it can be expanded with the Fourier series as $H\left(t\right)=\sum_{n=-\infty}^{\infty}H_{n}e^{in\omega t}$, where
\begin{small}
\begin{eqnarray}
H_{n} & = & \omega_{0}J_{z}\delta_{n,0}\!+\!\frac{1}{2}\sum_{i=x,y,z}A_{i}\left(e^{i\phi_{i}}\delta_{n,n_{i}}\!+ \! e^{-i\phi_{i}}\delta_{n,-n_{i}}\right)J_{i}.
\end{eqnarray}
\end{small}

Then the Floquet Hamiltonian in the frequency space as $H_{F}$ can be built with $H_{n}$, and the element in the row $n'$ and column $n$ of the Floquet Hamiltonian is written as~\cite{Floquet1inPhysRev.138.B979,Floquet1998byGRIFONI1998229,Floquet1byCHU20041}
\begin{equation}
\left(H_{F}\right)_{n}^{n'}=H_{n'-n}+\delta_{n',n}n\omega E,\label{eq:Floquet Hamiltonian}
\end{equation}
where $E$ is the identical matrix.
In this way, the quasienergy and the Floquet states of the periodical system, i.e. the eigenvalues $\varepsilon_{\alpha}$ and the corresponding eigenstates $|\Phi_\alpha\rangle$ of $H_{F}$ can be obtained. Furthermore, the final state of our QBs system is given by Floquet theorem as~(see Appendix B for details)~\cite{Floquet1inPhysRev.138.B979,Floquet1byCHU20041,Floquet2009inPhysRevA.79.032301}
\begin{small}
\begin{eqnarray}
|\Psi\left(t\right)\rangle = \sum_{\alpha=1}^{N+1}\sum_{n,n'=-\infty}^{\infty}e^{i\left(n-n'\right)\omega t}e^{-i\varepsilon_{\alpha}t}|\Phi_{\alpha}^{n}\rangle\langle\Phi_{\alpha}^{n'}|\frac{N}{2},-\frac{N}{2}\rangle,
\end{eqnarray}
\end{small}
where $\varepsilon_{\alpha},\alpha=1,2,\cdots,N+1$ are the $N+1$ eigenvalues of the Floquet Hamiltonian $H_{F}$ in one period of frequency space such as $\varepsilon_{\alpha}\in[-\frac{\omega}{2}+k\omega,\frac{\omega}{2}+k\omega),k\in\mathbb{Z}$,
and $|\Phi_{\alpha}^{n}\rangle$ is the element in corresponding eigenstate $|\Phi_{\alpha}\rangle=\left(\begin{array}{cccc}
\cdots, & |\Phi_{\alpha}^{n_{s}}\rangle, & |\Phi_{\alpha}^{n_{s}+1}\rangle, & \cdots
\end{array}\right)^T$.

\begin{figure}[ht]
	\centering 
	\includegraphics[width=90mm]{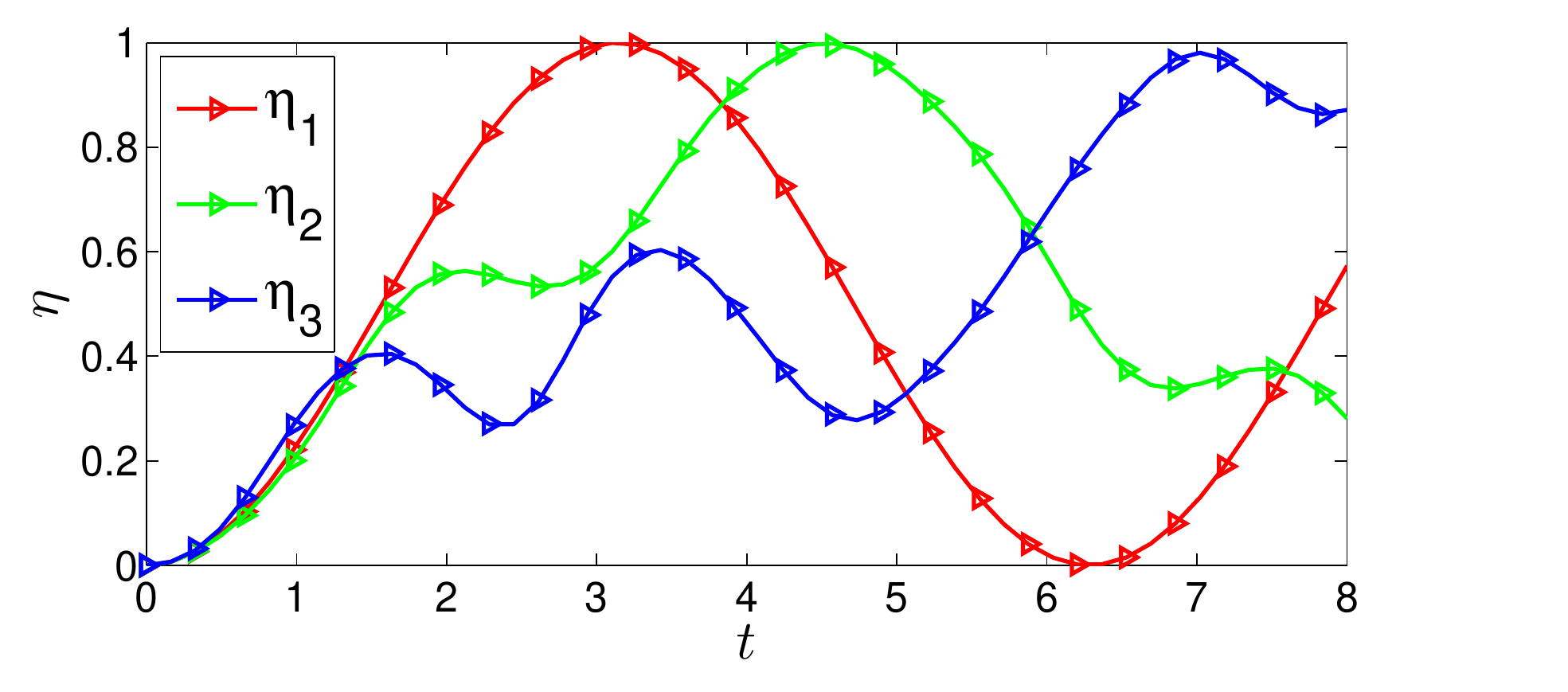}
	\protect
	\caption{The solid line and the triangles stand for the numerical results and the
		Floquet theorem's results of charge saturation, respectively.
		$\eta_{1}$ (marked in red) is obtained by setting $A_{x,y}=\omega_{0}$,
		$A_{z}=0$, $\omega_{x,y}=\omega$, $\phi_{x}=0$, $\phi_{y}=-\frac{\pi}{2}$.
		$\eta_{2}$ (marked in green) is obtained by setting $A_{x,y}=\omega_{0}$, $A_{z}=2\omega_{0}$, $\omega_{x,y}=\omega$,
		$\omega_{z}=2\omega$, $\phi_{x}=0$, $\phi_{y}=-\frac{\pi}{2}$,
		$\phi_{z}=\pi$, and $\eta_{3}$ (marked in blue) is obtained by setting
		$A_{x,y,z}=\omega_{0}$, $\omega_{x}=\omega$, $\omega_{y}=2\omega$, $\omega_{z}=3\omega$, $\phi_{x}=0$, $\phi_{y}=-\frac{\pi}{2}$,
		$\phi_{z}=\frac{3}{2}\pi$.}
	
	\label{fig:Fig4}
\end{figure}

So with the final state, the charge saturation is written as
\begin{scriptsize}
\begin{eqnarray}
\eta\left(t\right) \!= \!  \frac{1}{N}\sum_{\alpha,\tilde{\alpha}=1}^{N+1}\sum_{n,n',\tilde{n},\tilde{n}'=-\infty}^{\infty} \! \! \! \! \! \! \! \! \! \! \! \! \!\! \! \! \! \! \! \! e^{-i\tilde{\varepsilon}_{\alpha,\tilde{\alpha}}^{n,n',\tilde{n},\tilde{n}'}t}\left(\Phi_{\alpha,N}^{n'}\right)^{*}\Phi_{\tilde{\alpha},N}^{\tilde{n}'}\langle\Phi_{\tilde{\alpha}}^{\tilde{n}}|J_{z}|\Phi_{\alpha}^{n}\rangle+\frac{1}{2},\label{eq:eta_Floquet}
\end{eqnarray}
\end{scriptsize}
where ~$\tilde{\varepsilon}_{\alpha,\tilde{\alpha}}^{n,n',\tilde{n},\tilde{n}'}=\varepsilon_{\alpha}-\varepsilon_{\tilde{\alpha}}+\left(\tilde{n}-\tilde{n}'+n-n'\right)\omega$~ is the difference between the quasienergy of different period and ~$\Phi_{\alpha,N}^{n'}=\langle\frac{N}{2},-\frac{N}{2}|\Phi_{\alpha}^{n'}\rangle$~
is the last element of $|\Phi_{\alpha}^{n'}\rangle$. We give some examples of the results obtained by Floquet theorem in Fig.\ref{fig:Fig4}, which agree with the exact numerical results well. 

Moreover, when we set $N=1$ and $t=kT, k\in \mathbb{Z}$ to analyze the charging process of single battery unit, the charge saturation can be simplified as
\begin{small}
\begin{eqnarray}
\eta\left(kT\right) & = & \sum_{\alpha,\tilde{\alpha}=1}^{2}C_{\alpha,\tilde{\alpha}}e^{-i k\left(\varepsilon_{\alpha}-\varepsilon_{\tilde{\alpha}}\right)T}+\frac{1}{2}\nonumber \\
& = & 2\left|C_{1,2}\right|\cos\left[k \Delta\varepsilon T - \arg\left(C_{1,2}\right)\right]+C_{1,1}+C_{2,2}+\frac{1}{2},\label{eq:etaT}
\end{eqnarray}
\end{small}
where $C_{\alpha,\tilde{\alpha}}=\sum_{n,n',\tilde{n},\tilde{n}'=-\infty}^{\infty}\left(\Phi_{\alpha,N}^{n'}\right)^{*}\Phi_{\tilde{\alpha},N}^{\tilde{n}'}\langle\Phi_{\tilde{\alpha}}^{\tilde{n}}|\sigma_{z}|\Phi_{\alpha}^{n}\rangle$,
and $\Delta\varepsilon=\varepsilon_{2}-\varepsilon_{1}<\omega$ is
the difference of quasienergy in one period. Eq.~(\ref{eq:etaT}) tells that when $t=kT$, the charge saturation $\eta\left(t\right)$ of QB unit locate at the cosine curve which is determined by the quasienergy and the Floquet states.

\section{Further analysis with some numerical results}\label{sec4}

After the analytical analysis of the charging system, we further explore the relationship between the parameters and the time-dependent charge saturation with some numerical results.

\subsection{Numerical results when adjusting the charging strength and distributions of charging strength in vertical directions}\label{sec4-1}
\begin{figure}[ht]
	\centering 
	\includegraphics[width=90mm]{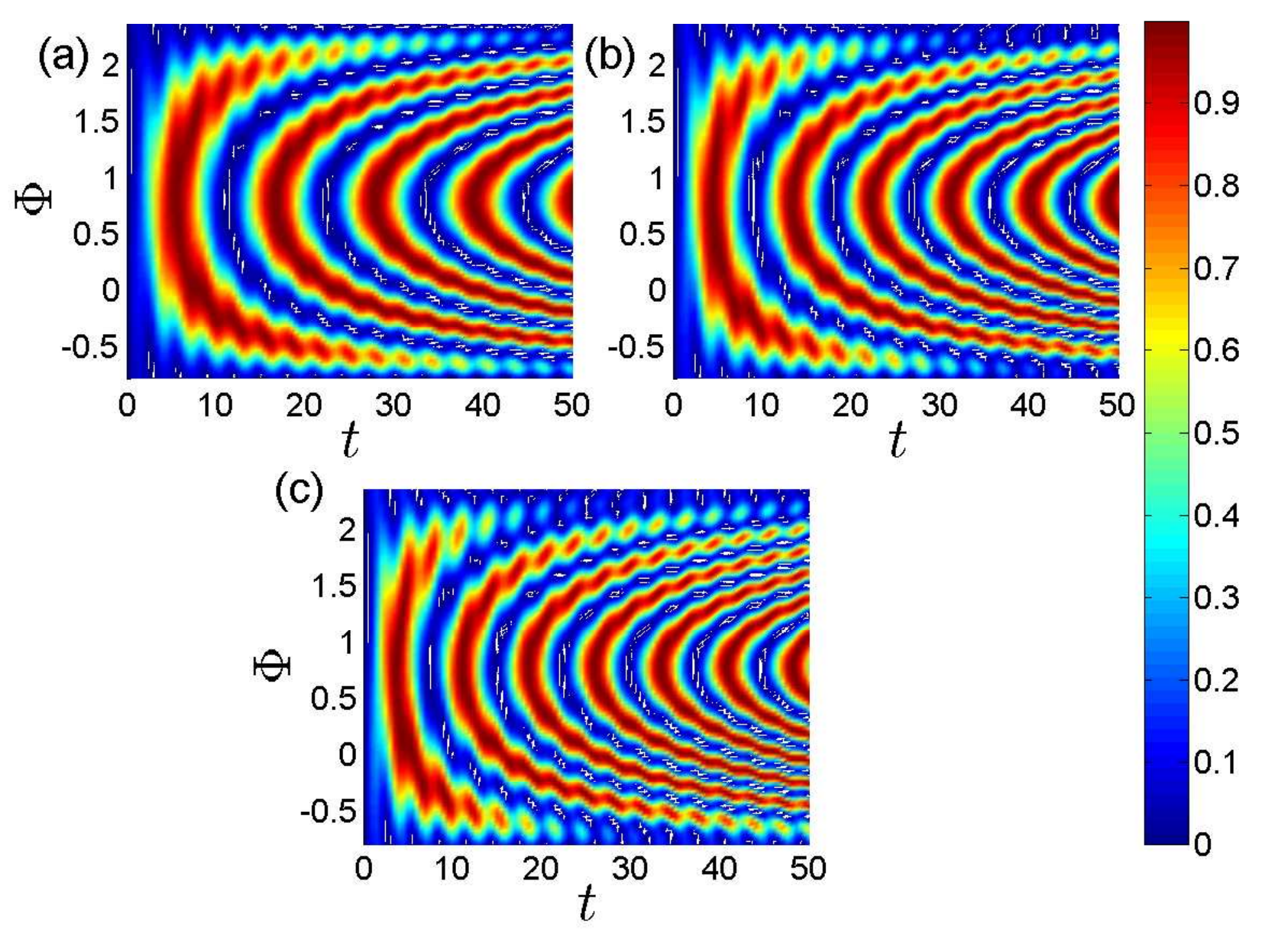}
	\protect\caption{The numerical results of charge saturation with the variety of $\Phi$ as $\eta\left(\Phi,t\right)$ when setting $\Theta=0,$ $\omega_{x}=\omega_{y}=\omega,$ and $\phi_{x}=0,$ $\phi_{y}=-\frac{\pi}{2}$, where the three subfigures (a), (b) and (c) stand for the results by setting  $\omega/\omega_{0}=1$ and $A/\omega_{0}=0.8,$
		$1$, $1.2$, respectively. }
	\label{fig:Fig5}
\end{figure}

After comparing the $H$-system and the $\tilde{H}$-system, we extend the analysis to the Hamiltonian with a more general charging field by setting $\Theta=0,\omega_{x}=\omega_{y}=\omega,\phi_{x}=0,\phi_{y}=-\frac{\pi}{2}$,
as
\begin{eqnarray}
H\left(t\right) & = & \omega_{0}J_{z}+A\left[\cos\Phi\cos\left(\omega t\right)J_{x}+\sin\Phi\sin\left(\omega t\right)J_{y}\right],\label{eq:H_Phi}
\end{eqnarray}
which contains both the $H$-system and the $\tilde{H}$-system. Then we give numerical results of the charge saturation as $\eta\left(\Phi,t\right)$ in Fig.\ref{fig:Fig5} to show the relationship between the charge saturation and the variety of charging strength and the distributions of charging strength. The range of $\Phi$ is moved from $\left[0,\pi\right]$ to $\left[-\frac{1}{4}\pi,\frac{3}{4}\pi\right]$ here to show the symmetry of the results.

In the $H$-system and the $\tilde{H}$-system, the charging time to reach the saturated mode has been proved to be inversely proportional to the charging strength $A$. And the results in Fig.\ref{fig:Fig5}(a)(b)(c) with different charging strength also tell that the charging time is shortened when improving the charging strength. What is more, we can get that the charging power reach the maximum by setting $\Phi=\pi/4$, which means that the $H$-system is the most effective charging system when adjusting the distribution of the charging strength.

What is more, when setting $\Phi=-\frac{1}{4}\pi+k\pi$, the Hamiltonian turns into $H=\omega_{0}J_{z}+\frac{\sqrt{2}}{2}A\left[\cos\left(\omega t\right)J_{x}-\sin\left(\omega t\right)J_{y}\right]$,
which can be obtained by flipping the driving in $y$ direction from $H$-system. The numerical results show that the QBs can hardly be charged in this situation, which is useless for our QBs charging system. But it provides a solution that we can make a shield for the quantum system from the influence of the harmonic field in vertical direction by introducing another harmonic driving field as it is shown in this situation.

\subsection{Numerical results when adding charging field in parallel direction on the $H$-system }\label{sec4-2}

Now we take the $H$-system as a basic system, and add a general charging field in parallel direction as $A\sin\Theta\cos\left(\omega_{z}t+\phi_{z}\right)J_{z}$. So the Hamiltonian of the charging system turns to be
\begin{eqnarray}
H\left(t\right) & = & \left[\omega_{0}+A\sin\Theta\cos\left(\omega_{z}t+\phi_{z}\right)\right]J_{z}\nonumber\\
& &+\frac{\sqrt{2}}{2}A\cos\Theta\left[\cos\left(\omega t\right)J_{x}+\sin\left(\omega t\right)J_{y}\right].\label{eq:H_Theta}
\end{eqnarray}

\begin{figure}[ht]
\centering 
\includegraphics[width=90mm]{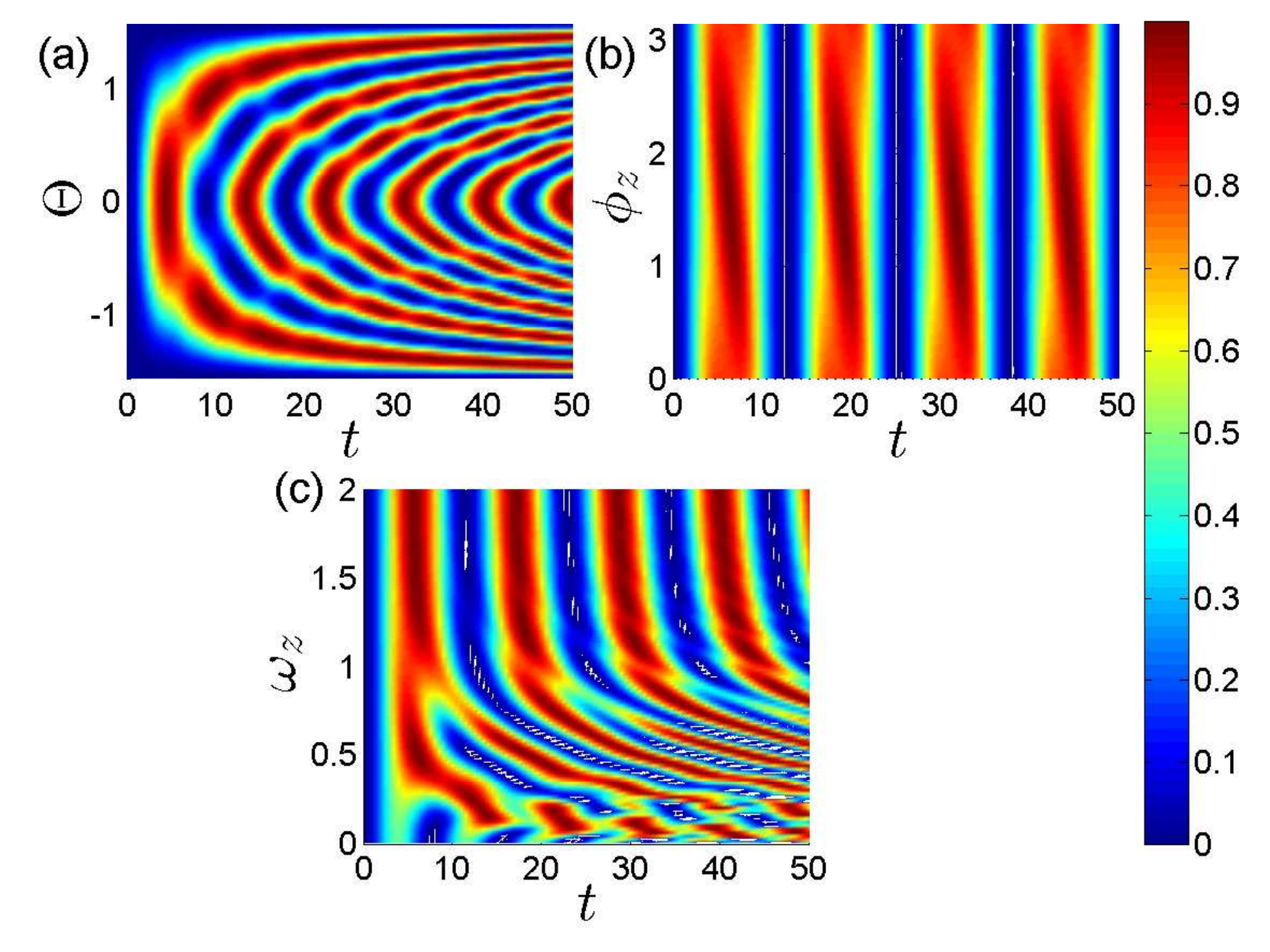}
\protect\caption{(a) The time-dependent charge saturation with the variety of $\Theta$
as $\eta\left(\Theta,t\right)$ when setting $\omega_{z}=\omega$ and $\phi_{z}=0$ in Eq.~(\ref{eq:H_Theta}).
(b) The time-dependent charge saturation with the variety of $\phi_{z}$
as $\eta\left(\phi_{z},t\right)$ when setting $\Theta=\arccos\left(0.8\right)$ and $\omega_{z}=\omega$ in Eq.~(\ref{eq:H_Theta}). (c) The time-dependent charge saturation
with the variety of $\omega_{z}$ as $\eta\left(\omega_{z},t\right)$
when setting $\Theta=\arccos\left(0.8\right)$ and $\phi_{z}=0$ in Eq.~(\ref{eq:H_Theta}).}

\label{fig:Fig6}
\end{figure}

At first, we set $\omega_{z}=\omega$, $\phi=0$, and get the numerical results as $\eta\left(\Theta,t\right)$ in Fig.\ref{fig:Fig6}(a) to describe the relationship between the charge saturation and the distribution of charging power. The results tell that the charge saturation will decrease when we distribute harmonic power from the vertical direction into the parallel direction.

Then for further analysis, we set $\Theta=\arccos\left(0.8\right)$ to get a system with $A=1$, and $A_{x}=A_{y}=0.8$, which is easy to compare with the results in the Fig.\ref{fig:Fig5}(a) and Fig.\ref{fig:Fig5}(b). And we calculate the numerical results of saturation with the variety of $\phi_{z}$ and $\omega_{z}$ numerically, and show them in Fig.\ref{fig:Fig6}(a) and Fig.\ref{fig:Fig6}(b), respectively. The results tell that the charging power can be enhanced with proper setting of $\phi_{z}$ and $\omega_{z}$. But the comparison of the results with the Fig.\ref{fig:Fig5}(a) and Fig.\ref{fig:Fig5}(b) shows that the $H$-system is still more effective, which means that arbitrary driving in the parallel direction is negative for our charging system.

\begin{figure}[ht]
\centering 
\includegraphics[width=90mm]{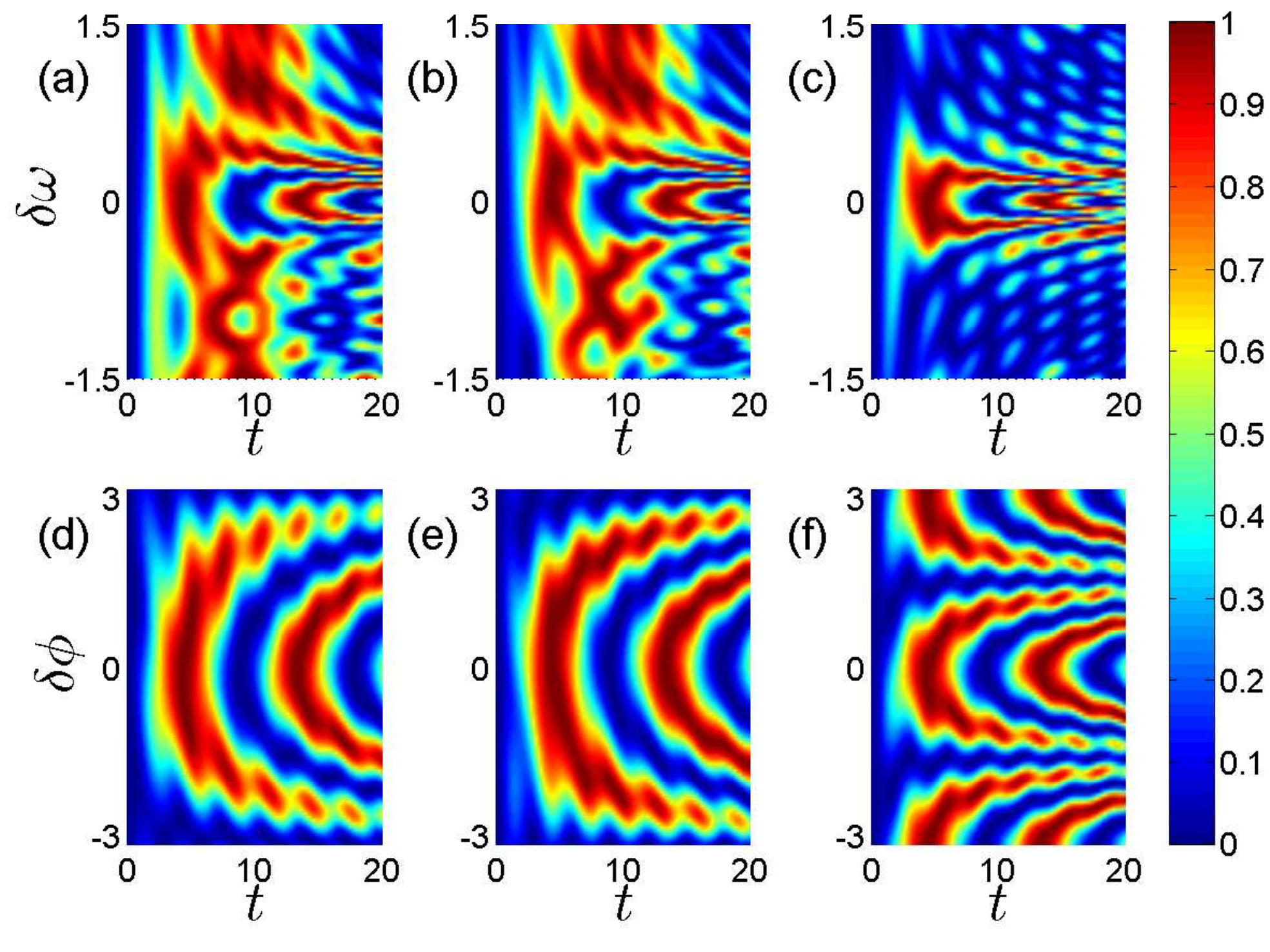}
\protect\caption{(a) The charge saturation with perturbation of frequency in $x$-direction as Eq.~(\ref{eq:4-3-1});
(b) The charge saturation with perturbation of frequency in $y$-direction as Eq.~(\ref{eq:4-3-2});
(c) The charge saturation with opposite perturbations of frequency in $x$-direction and $y$-direction as Eq.~(\ref{eq:4-3-3});
(d) The charge saturation with perturbation of initial phase in $x$-direction as Eq.~(\ref{eq:4-3-4});
(e) The charging saturation with perturbation of initial phase in $y$-direction as Eq.~(\ref{eq:4-3-5});
(f) The charge saturation with opposite perturbations of initial phase in $x$-direction and $y$-direction as Eq.~(\ref{eq:4-3-6}). }
\label{fig:Fig7}
\end{figure}

\subsection{Numerical results with perturbation of frequency and phase on the $H$-system}\label{sec4-3}

It is shown that the $H$-system is the optimal charging system with the former results. Then we change the driving frequency and the initial phase of the two vertical directions as the perturbations on the $H$-system to find if the charging efficiency could be improved. The Hamiltonian with different perturbations are set as
\begin{scriptsize}
\begin{eqnarray}
H_{1}\left(\delta\omega,t\right) & = & \omega_{0}J_{z}+\frac{\sqrt{2}}{2}A\left[\cos\left[\left(\omega+\delta\omega\right)t\right]J_{x}+\sin\left[\left(\omega\right)t\right]J_{y}\right],\label{eq:4-3-1}\\
H_{2}\left(\delta\omega,t\right) & = & \omega_{0}J_{z}+\frac{\sqrt{2}}{2}A\left[\cos\left[\left(\omega\right)t\right]J_{x}+\sin\left[\left(\omega+\delta\omega\right)t\right]J_{y}\right],\label{eq:4-3-2}\\
H_{3}\left(\delta\omega,t\right) & = & \omega_{0}J_{z}+\frac{\sqrt{2}}{2}A\left[\cos\left[\left(\omega\!+\!\delta\omega\right)t\right]J_{x}\!+\!\sin\left[\left(\omega-\delta\omega\right)t\right]J_{y}\right],\label{eq:4-3-3}\\
H_{1}\left(\delta\phi,t\right) & = & \omega_{0}J_{z}+\frac{\sqrt{2}}{2}A\left[\cos\left(\omega t+\delta\phi\right)J_{x}+\sin\left(\omega t\right)J_{y}\right],\label{eq:4-3-4}\\
H_{2}\left(\delta\phi,t\right) & = & \omega_{0}J_{z}+\frac{\sqrt{2}}{2}A\left[\cos\left(\omega t\right)J_{x}+\sin\left(\omega t+\delta\phi\right)J_{y}\right],\label{eq:4-3-5}\\
H_{3}\left(\delta\phi,t\right) & = & \omega_{0}J_{z}+\frac{\sqrt{2}}{2}A\left[\cos\left(\omega t+\delta\phi\right)J_{x}+\sin\left(\omega t-\delta\phi\right)J_{y}\right].\label{eq:4-3-6}
\end{eqnarray}
\end{scriptsize}

The numerical results, as $\eta_{i}\left(\delta\omega,t\right)$ and $\eta_{i}\left(\delta\phi,t\right),i=1,2,3$, are given in Fig.\ref{fig:Fig7}, which stand for the time-dependent charge saturation with perturbation of frequency or phase in one or two directions, respectively.

The results tell that the $H$-system is optimal for saturated charging mode. But for the unsaturated charging mode, the charging efficiency can be enhanced with some proper adjustment of parameters. For instance, when the effective charge saturation is set as $\eta=0.4$, it is obvious that the charging time can be shortened by adding a proper negative perturbation of frequency in Eq.~(\ref{eq:4-3-3}) as Fig.\ref{fig:Fig7}(c) shows, and so can it by adding a proper positive perturbation of phase in Eq.~(\ref{eq:4-3-5}) or a proper negative perturbation of phase in Eq.~(\ref{eq:4-3-6}) as the Fig.\ref{fig:Fig7}(e) and Fig.\ref{fig:Fig7}(f) shows, respectively.

\section{Conclusion}\label{sec5}

To conclude, in this work, we build the QBs system with $N$ two-level atoms and charge it with a controlled general harmonic driving field. According to the analytical and numerical analysis of our QBs charging system, we find out that the QBs can be fully charged when the driving field of charging system is set with proper parameters. And with the same charging strength $A$, when the other parameters of the driving field are set as $\Theta=0, \Phi=\frac{\pi}{4}$, and $\Delta \phi=\phi_{x}-\phi_{y}=\frac{\pi}{2}$, the QBs are fully charged in the shortest time. It is the optimal charging system for the saturated charging mode.

What is more, the results in Sec.~\ref{sec4-3} tell that when considering the unsaturated charging mode, the charging efficiency can be further enhanced by introducing proper perturbation of frequency or phase on the charging system, as it is shown in Fig.\ref{fig:Fig7}.

We believe that in the near future, QBs, as a kind of important quantum device, will step from theory into the practice. The charging model of QBs we studied in this paper is supposed to be realized physically, for instance, we can take the solid-state platform to build our battery and charge it with Raman laser beams as the driving field~\cite{QB2018inPhysRevLett.120.117702}. It can be seen that the bidirectional and biphase charging mode as $H$-system and the unsaturated charging mode proposed in this paper will provide reasonable reference for many kinds of schemes which are aimed at improving the charging efficiency of this kind of QBs.

\section*{Appendix}

\setcounter{equation}{0}
\setcounter{subsection}{0}
\renewcommand{\theequation}{A\arabic{equation}}
\renewcommand{\thesubsection}{A\arabic{subsection}}

\subsection*{\textbf{A. Unitary transformation in su(2) system}}\label{Appendix A}
Arbitrary unitary transformation operator in su(2) system can be constructed
as $U=e^{iA}$ with $A=\bm{a}\cdot\bm{J}$, where $\bm{J}=\left(J_{x},J_{y},J_{z}\right)$
is the generator of su(2) algebra. With the commutation relation for
su(2) algebra as $\left[J_{i},J_{j}\right]=i\epsilon_{ijk}J_{k}$,
, the transformation for any operators in the space can be written
as
\begin{small}
\begin{eqnarray}
e^{iA}Be^{-iA} & = & \sum_{n=0}^{\infty}\frac{1}{n!}\left(iA^{\times}\right)^{n}B=\sum_{n=0}^{\infty}\frac{\left(-1\right)^{n}}{n!}\left(\bm{a}\times\right)^{n}\bm{b}\cdot\bm{J} \\
 & = & \bm{b}\cdot\bm{J}\!-\!\frac{\sin a}{a}\bm{a}\times\bm{b}\cdot\bm{J}\!-\!\frac{1}{a^{2}}\left(\cos a \!-\!1\right)\left(\bm{a}\times\right)^{2}\bm{b}\!\cdot\!\bm{J},\nonumber
\end{eqnarray}
\end{small}where $B=\bm{b}\cdot\bm{J}$ and $a=|\bm{a}|$, $b=|\bm{b}|$. Then
if $\bm{a}\perp\bm{b}$, the result can be simplified as
\begin{eqnarray}
e^{iA}Be^{-iA} & = & \cos a\left(\bm{b}\cdot\bm{J}\right)-\frac{\sin a}{a}\left(\bm{a}\times\bm{b}\cdot\bm{J}\right) \nonumber\\
 & = & b\left[\cos a\bm{e}_{b}-\sin a\left(\bm{e}_{a}\times\bm{e}_{b}\right)\right]\cdot\bm{J}.
\end{eqnarray}

\subsection*{\textbf{B. Floquet theorem in quantum system}}\label{Appendix B}

\setcounter{equation}{0}
\setcounter{subsection}{0}
\renewcommand{\theequation}{B\arabic{equation}}
\renewcommand{\thesubsection}{B\arabic{subsection}}
For a quantum system of $N$ dimensions, the Schr\"{o}dinger equation
$i\partial_{t}|\Psi\left(t\right)\rangle=H\left(t\right)|\Psi\left(t\right)\rangle$
has $N$ linearly independent solutions as $|\Psi_{\alpha}\left(t\right)\rangle,\alpha=1,\cdots,N$.
And the wave function of system is represented as $|\Psi\left(t\right)\rangle=\sum_{\alpha=1}^{N}c_{\alpha}|\Psi_{\alpha}\left(t\right)\rangle$,
where $\left\{ c_{\alpha}\right\} $ is determined by the initial
state $|\Psi\left(t\right)\rangle$. When the Hamiltonian of the system
is periodic as $H\left(t+\frac{2\pi}{\omega}\right)=H\left(t\right)$,
the Floquet theorem tells that we can find a real number $\varepsilon_{\alpha}$
and a periodic wave function $|\Phi_{\alpha}\left(t\right)\rangle$
with the same period as $T=\frac{2\pi}{\omega}$ to rewrite $|\Psi_{\alpha}\left(t\right)\rangle$
as~\cite{Floquet1inPhysRev.138.B979,Floquet1998byGRIFONI1998229,Floquet1byCHU20041}
\begin{equation}
|\Psi_{\alpha}\left(t\right)\rangle=e^{-i\varepsilon_{\alpha}t}|\Phi_{\alpha}\left(t\right)\rangle.\label{eq:Psialphat}
\end{equation}

Then substituting the expression into Schr\"odinger equation, we can
get the equation of $|\Phi_{\alpha}\left(t\right)\rangle$ as
\begin{equation}
\left[H\left(t\right)-i\partial_{t}\right]|\Phi_{\alpha}\left(t\right)\rangle=\varepsilon_{\alpha}|\Phi_{\alpha}\left(t\right)\rangle.\label{eq:PhiEquation}
\end{equation}
Furthermore, we expand the $H\left(t\right)$ and $|\Phi_{\alpha}\left(t\right)\rangle$
with the frequency $\omega$ into Fourier series as
\begin{eqnarray}
H\left(t\right) & = & \sum_{n=-\infty}^{\infty}H_{n}e^{in\omega t},\\
|\Phi_{\alpha}\left(t\right)\rangle & = & \sum_{n=-\infty}^{\infty}|\Phi_{\alpha}^{n}\rangle e^{in\omega t},
\end{eqnarray}
and substitute them into Eq.~(\ref{eq:PhiEquation}). In this way, we can get
\begin{eqnarray}
& &\sum_{n'=-\infty}^{\infty}\sum_{n=-\infty}^{\infty}H_{n'}e^{i\left(n'+n\right)\omega t}|\Phi_{\alpha}^{n}\rangle+\sum_{n=-\infty}^{\infty}n\omega|\Phi_{\alpha}^{n}\rangle e^{in\omega t}\nonumber \\
& = &\varepsilon_{\alpha}\sum_{n=-\infty}^{\infty}|\Phi_{\alpha}^{n}\rangle e^{in\omega t}.
\end{eqnarray}
Then with  $\frac{1}{T}\int_{0}^{T}dt\left(e^{-ik\omega t}e^{in\omega t}\right)=\delta_{k,n}$, this time-dependent equation for $\left\{ |\Phi_{\alpha}^{n}\rangle\right\}$ can be transferred into a time-independent one as
\begin{eqnarray}
\sum_{n=-\infty}^{\infty}\left(H_{k-n}+\delta_{k,n}n\omega\right)|\Phi_{\alpha}^{n}\rangle & = & \varepsilon_{\alpha}|\Phi_{\alpha}^{k}\rangle,\label{eq:Floquet Equation0}
\end{eqnarray}
for any $k\in \mathbb{Z}$. By introducing the Floquet Hamiltonian
$H_{F}$ defined by Eq.~(\ref{eq:Floquet Hamiltonian}), the
matrix form of Eq.~(\ref{eq:Floquet Equation0}) can be written as
\begin{small}
\begin{eqnarray}
\left(\begin{array}{cccc}
\ddots & \vdots & \vdots & \iddots\\
\cdots & \left(H_{F}\right)^{k_{s}}_{n_{s}} & \left(H_{F}\right)^{k_{s}}_{n_{s}+1} & \cdots\\
\cdots & \left(H_{F}\right)^{k_{s}+1}_{n_{s}} & \left(H_{F}\right)^{k_{s}+1}_{n_{s}+1} & \cdots\\
\iddots & \vdots & \vdots & \ddots
\end{array}\right)
\left(\begin{array}{c}
\vdots\\
|\Phi_{\alpha}^{n_{s}}\rangle\\
|\Phi_{\alpha}^{n_{s}+1}\rangle\\
\vdots
\end{array}\right)
  =  \varepsilon_{\alpha}
\left(\begin{array}{c}
 \vdots\\
|\Phi_{\alpha}^{k_{s}}\rangle\\
|\Phi_{\alpha}^{k_{s}+1}\rangle\\
\vdots
\end{array}\right),\label{eq:Floquet Equation1}
\end{eqnarray}
\end{small}
where $k_{s}$ and $n_{s}$ are indexes which are introduced to describe
the matrix. And Eq.~(\ref{eq:Floquet Equation1}) tells that $\varepsilon_{\alpha}$
and $|\Phi\rangle=\left(\begin{array}{c}\vdots\\
|\Phi_{\alpha}^{n_{s}}\rangle\\
|\Phi_{\alpha}^{n_{s}+1}\rangle\\
\vdots
\end{array}\right)=\left(\begin{array}{c}\vdots\\
|\Phi_{\alpha}^{k_{s}}\rangle\\
|\Phi_{\alpha}^{k_{s}+1}\rangle\\
\vdots
\end{array}\right)$ are eigenvalues and eigenstates of $H_{F}$, respectively. It is obvious that Eq.~(\ref{eq:Floquet Equation1}) has infinite solutions
but we only need $N$ of them. However, the expanding of solutions
is introduced by the periodic relationships of $\varepsilon_{\alpha}$
and $|\Phi{}_{\alpha}\left(t\right)\rangle$ as $ \left\{
\begin{array}{ccl}
\varepsilon_{\alpha'} & = & \varepsilon_{\alpha}+m\omega\\
|\Phi'_{\alpha}\left(t\right)\rangle & = & e^{im\omega t}|\Phi{}_{\alpha}\left(t\right)\rangle
\end{array} \right. $. So we can get the $N$ solutions of $\varepsilon_{\alpha}$
and $\left(\begin{array}{c}\vdots\\
|\Phi_{\alpha}^{n_{s}}\rangle\\
|\Phi_{\alpha}^{n_{s}+1}\rangle\\
\vdots
\end{array}\right)$, with $\alpha=1,\cdots,N$ by setting $\varepsilon_{\alpha}$ in one period of frequency space, such as $\varepsilon_{\alpha}\in[-\frac{\omega}{2},\frac{\omega}{2})$.

With the results of $\left\{ \varepsilon_{\alpha}\right\} $ and $\left\{ |\Phi_{\alpha}^{n}\rangle\right\} $,
the wave function can be calculated as $|\Psi\left(t\right)\rangle=\sum_{\alpha,n}c_{\alpha}|\Phi_{\alpha}^{n}\rangle e^{in\omega t}e^{-i\varepsilon_{\alpha}t}$.
And by setting $t=0$, we get $|\Psi\left(0\right)\rangle=\sum_{\alpha,n}c_{\alpha}|\Phi_{\alpha}^{n}\rangle=\left(\sum_{n}|\Phi_{1}^{n}\rangle,\sum_{n}|\Phi_{2}^{n}\rangle,\cdots,\sum_{n}|\Phi_{N}^{n}\rangle\right)
\left(\begin{array}{c}
c_{1}\\
c_{2}\\
\vdots\\
c_{N}
\end{array}\right)$. In this way, the series of 
coefficients $\left\{ c_{\alpha}\right\} $
can be expressed as
\begin{equation}
\left(\begin{array}{c}
c_{1}\\
c_{2}\\
\vdots\\
c_{N}
\end{array}\right)=\left(\sum_{n}|\Phi_{1}^{n}\rangle,\sum_{n}|\Phi_{2}^{n}\rangle,\cdots,\sum_{n}|\Phi_{N}^{n}\rangle\right)^{-1}|\Psi\left(0\right)\rangle.\label{eq:c_alpha}
\end{equation}
Then by substituting the coefficients into the expression of wave
function, we can get the final state as $|\Psi\left(t\right)\rangle=U\left(t\right)|\Psi\left(0\right)\rangle$,
where
\begin{eqnarray}
U\left(t\right) & = & \sum_{\alpha=1}^{N}\sum_{n,n'=-\infty}^{\infty}|\Phi_{\alpha}^{n}\rangle\langle\Phi_{\alpha}^{n'}|e^{i\left(n-n'\right)\omega t}e^{-i\varepsilon_{\alpha}t}.\label{eq:Ut_Floquet}
\end{eqnarray}
Then with the initial state of our system as $|\Psi\left(0\right)\rangle=|\frac{N}{2},-\frac{N}{2}\rangle$,
we can get the charge saturation as Eq.~(\ref{eq:eta_Floquet}).

\section*{Acknowledgements}

We acknowledge helpful discussions with Yanming Che, Yuguo Su and Gangcheng Wang. This work was supported by the National Key Research and Development Program of China (Grants No.~2017YFA0304202 and No.~2017YFA0205700), the NSFC through Grant No.~11875231 and No.~11935012, and the Fundamental Research Funds for the Central Universities through Grant No.~2018FZA3005.

\section*{Conflict of Interest}

The authors declare no conflict of interest.

\section*{Keywords}

quantum battery, driven system, harmonic driving, charge saturation, su(2), Floquet theorem, unsaturated charging mode

%



\end{document}